# Scalable high-mobility graphene/hBN heterostructures

*Leonardo Martini[1,*], Vaidotas Miseikis[1,2], David Esteban[3], Jon Azpeitia[3], Sergio Pezzini[4], Paolo Paletti[1,2], Michal W. Ochapski[1,2], Domenica Convertino[1], Mar Garcia Hernandez[3], Ignacio Jimenez[3], Camilla Coletti[1,2,*]*

1. Center for Nanotechnology Innovation @NEST, Istituto Italiano di Tecnologia, Piazza San Silvestro 12, 56127, Pisa, Italy
2. Graphene Labs, Istituto Italiano di Tecnologia, Via Morego 30, I-16163 Genova, Italy
3. Instituto de Ciencia de Materiales de Madrid, Consejo Superior de Investigaciones Científicas, E-28049 Madrid, Spain
4. NEST, Istituto Nanoscienze-CNR and Scuola Normale Superiore, Piazza San Silvestro 12, 56127, Pisa, Italy

*leonardo.martini@iit.it, camilla.coletti@iit.it*

Graphene-hexagonal boron nitride (hBN) scalable heterostructures are pivotal for the development of graphene-based high-tech applications. In this work we demonstrate the realization of high-quality graphene-hBN heterostructures entirely obtained with scalable approaches. hBN continuous films were grown via ion beam assisted physical vapor deposition (IBAD-PVD) directly on commercially-available $SiO_2$/Si, and used as receiving substrates for graphene single-crystal matrixes grown by chemical vapor deposition (CVD) on copper. The structural, chemical and electronic properties of the heterostructure were investigated by atomic force microscopy (AFM), Raman spectroscopy and electrical transport measurements. We demonstrate graphene carrier mobilities exceeding 10000 $cm^2$/Vs in ambient conditions, 30% higher than those directly measured on $SiO_2$/Si. We prove the scalability of our approach by measuring more than 100 transfer length method (TLM) devices over centimeter scale, which present an average carrier mobility of 7500 ± 850 $cm^2$/Vs. The reported high-quality all-scalable heterostructures are of relevance for the development of graphene-based high-performing electronic and optoelectronic applications.

## Introduction
In recent years hexagonal boron nitride (hBN) has attracted attention as a promising encapsulant for graphene[1][2] and other 2D-materials[3], due to its remarkable structural, chemical and electronic properties. Like graphene, hBN is a layered material with a hexagonal lattice, it can be conveniently obtained via mechanical exfoliation from bulk crystals, and presents high chemical stability. Thanks to the small (~1.8%) difference in lattice parameters between graphene and hBN[4] and its atomically flat surface, hBN can be integrated into graphene-based heterostructures with an effective minimization of extrinsic disorder[5]. Moreover, hBN presents a bandgap as large as 6 eV[6][7], a dielectric constant of 3.4[8] comparable with that of silicon dioxide ($SiO_2$) and a very high breakdown voltage (i.e. 21 MV/cm [9]), which make it a suitable dielectric for the realization of field-effect transistor (FET) devices. When used to encapsulate graphene and other two-dimensional (2D) materials, hBN is effective in preserving the material quality and stability[1][10] and reducing the ambient induced contamination[11] with a beneficial effect on the electrical transport properties[12][1].

For most envisaged high-tech applications in the fields of photonics, optoelectronics and spintronics, hBN has soon become the ideal encapsulant material, capable of yielding graphene-based devices with the required performances[13][14]. Therefore, the scalable synthesis of hBN has become a crucial field of research. hBN thin films have been obtained via chemical vapor deposition (CVD) and molecular beam epitaxy (MBE) on several metallic substrates, such as copper[15][16], platinum[17], cobalt[18], and nickel[19]. Indeed, the CVD synthesis of monolayer and few-layer hBN is by now an established technique and the material is presently commercially available[20]. However, the synthesis of hBN films with a thickness of tens of nanometers, suitable to be adopted for bottom and top graphene encapsulation, as well as serving as a gate dielectric in electronic and photonic devices, is still considered a challenge. In first place, there is an objective difficulty in obtaining hBN whose quality matches that of exfoliated flakes from bulk crystals[21][1][22]. Also, although progresses have been reported for the CVD growth of thick hBN films both on metallic[23][24] and dielectric[25][26] substrates, there are significant challenges in identifying synthesis processes which comply with industrial requirements for CMOS integration such as metal contamination control (below $10^{10}$ atoms/cm$^2$) [14] [27]. The use of insulating substrates for the synthesis of hBN offers advantages such as the absence of metal contamination, though temperatures as high as 1400 °C [28] are often needed to obtain high quality hBN, which are not appealing from an industrial point of view. Other approaches have been explored to grow BN at low temperatures, such as microwave-assisted CVD (PECVD)[29] and plasma-enhanced atomic layer deposition (ALD)[30], but both those methods present safety limitations due to the use of toxic precursors as *n-Ethylmethylamine*[30]. The definition of a scalable and safe hBN growth approach yielding controlled thickness on insulating substrates would indeed be extremely attractive.

In this work, we report the realization of high-quality hBN/graphene heterostructures by employing scalable techniques which could be of potential interest for fab integration[33][33][34] . First, continuous films of nanocrystalline hBN with thicknesses of 10 nm are synthesized on $SiO_2$/Si at 1000 °C through a physical vapor deposition (PVD) approach, namely Ion Beam Assisted Deposition (IBAD). Subsequently, arrays of monolayer graphene single-crystals are grown via chemical vapor deposition (CVD) on copper and transferred with a semidry approach[33] on the target IBAD-hBN substrates.

Combined analyses of the spectroscopic, microscopic and transport properties of the heterostructure indicate that IBAD-hBN is a promising substrate for graphene devices as it provides a high-quality landscape for the graphene carriers. When measuring 109 devices the room temperature (RT) carrier mobility in graphene on IBAD-hBN is found to average at ~7500 $cm^2$/Vs, and the residual carrier density at the charge neutrality point is ~$2 \times 10^{11}$ $cm^{-2}$. As-processed devices initially show displacement of the Dirac point and gate hysteresis (attributed to the presence of trapped charges at the hBN/$SiO_2$ interface), which are both significantly reduced by vacuum treatment of the heterostructure.

## 1. Materials and Methods
### 1.1. hBN growth

Nanocrystalline hBN films were grown by IBAD on commercially available p-doped silicon substrates covered with 275 nm of thermally grown $SiO_2$, using nitrogen gas and solid boron as sources. The films can be grown with thicknesses ranging from 1 to 100 nm; in this work a thickness of 10 nm was used. The lateral size of the resulting hBN film is limited by the diameter of the ion gun, and in our setup homogenous films up to 3" wafers could be produced. Solid boron (Alfa-Aesar 12134) was evaporated using a 7 kV electron beam evaporator, while low energy nitrogen ions (average energy of 5 eV) were provided by a Kauffman ion gun fed with 5 standard cubic centimeters per minute (sccm) of high purity gas. The chamber base pressure was $10^{-7}$ mbar, reaching $10^{-4}$ mbar during the growth. The sample was maintained at 1000ºC during the growth. With the adopted growth technique it is possible to tune the properties of the material by changing the solid boron precursor from pure boron to boron-carbide ($B_4C$) thus yielding BN films (i.e., BNC) with a limited content of carbon (< 10%) and a different dielectric constant[31][34][35].

While both kinds of BN films (with and without carbon additive) were synthesized in this work, only pure hBN films were those ultimately adopted because of the higher crystallinity (see SI). Calibration of growth rates was done by contact profilometry, and the actual thickness was verified on test samples by UV-VIS spectrometry and spectroscopic ellipsometry. The quality and orientation of the adopted hBN films, which were found to exhibit a basal plane parallel to the substrate, was determined by X-ray Absorption Near Edge Structure (XANES)[36].

### 1.2. Graphene growth and transfer

Graphene single-crystal matrixes were grown by CVD in a deterministic pattern on electro-polished copper foils (Alpha-Aesar 99.8%) in a commercially-available cold-wall reactor (Aixtron 4" BM Pro), as reported in previous work[33]. Specifically, the substrate was first annealed in non-reducing argon atmosphere for 10 minutes, and the growth was then performed at 1060ºC with an argon flow of 900 sccm, 100 sccm of hydrogen and 1 sccm of methane, with a base pressure of 25 mbar. The graphene crystal arrays were transferred on the target substrates (i.e., $SiO_2$/Si with and without IBAD-hBN) through a deterministic semi-dry procedure[33]. The graphene on copper foil was covered with a double polymeric membrane of PMMA/PPC and baked at 90 °C[37], while a few millimetre-thick PDMS frame was applied on the edges of the sample, to ensure mechanical rigidity. The graphene was delaminated from the copper in a solution of 1 molar of NaOH [38][39] and transferred to the target substrate using a micromechanical stage to ensure the deterministic transfer. Once transferred, the polymer was removed by subsequent immersion in acetone and isopropanol. 2-step cleaning using remover AR 600-71 (Allresist) was performed to ensure the cleanliness of the graphene surface[40].

### 1.3. AFM and Raman characterization

Atomic force microscopy (AFM) was used to investigate the topography of the samples; it was performed using an Anasys AFM+ tool in non-contact mode and a Bruker Dimension Icon microscope used in ScanAsyst mode. AFM micrographs were analyzed using the software Gwyddion 2.54.

Raman spectroscopy was used to characterize the crystalline quality of both graphene and hBN. Raman data were acquired using a commercial Renishaw InVia spectrometer, with a laser wavelength of 532 nm. The Raman setup is linked to a microscope with mechanically controlled stage, thus allowing to perform spatially-resolved micro-Raman characterization with spot size in the order of 1 µm$^2$, defined by the 100x magnification lens. Raman characterization of hBN was performed using a laser power density of ~ 10 mW/µm$^2$. An acquisition time of 600 s was needed to detect representative hBN peaks[10]. Graphene was measured with a laser power density of 1.7 mW/µm$^2$. The statistics reported in the paper were obtained from spectra acquired on areas of 15x15 µm$^2$ with a step of 1 µm.

### 1.4. Device fabrication

Optical lithography and metal thermal evaporation (50 nm of gold on top of 5 nm of chromium) were used to pattern an array of markers on top of the hBN substrate, before the transfer of graphene. Hall-bar and Transfer Length Method (TLM) devices were fabricated using standard e-beam technique (EBL), with a Zeiss UltraPlus scanning electron microscope and Raith Multibeam lithography module. The graphene channels were defined with the first lithographic step. Reactive ion etching (RIE) with Ar/O$_2$ atmosphere for 45 seconds was used to remove graphene from the patterned areas. Subsequently, the metal contacts were defined via a second EBL step and thermal metal deposition of 50 nm of gold on top of 5 nm of chromium.

### 1.5. Electrical characterization

Electrical characterization was performed at room temperature and in air, using five micrometric positioners (MPI-corporation MP40) with 7 µm tungsten tips to provide signal and check the read-out. A Keithley 2450 sourcemeter, in high tension configuration, was used as DC source for the gate potential, with constant reading of the current to check for eventual leakages from the back gate. DC measurements, both in two and four-terminal configuration, were performed through a second Keithley 2450 sourcemeter. To assess the real graphene electrical performance, avoiding any contribution to the resistivity arising from the contacts, 4-probe measurement, both in Hall Bar devices and TLM were performed. AC measurements were carried out with a Signal Recovery 7260DSP in low frequency (10 – 100 Hz) configuration and differential voltage read-out. The constant-current was achieved using a large pre-resistor (4.7 MΩ) in series with the measured device. The orthogonal magnetic field in the Hall measurements (up to 1000 Oe) was provided through a commercial resistive electromagnet operated at room temperature.

## 2. Results and discussion

The morphology of the hBN film synthesized via IBAD was examined via AFM and compared to that of SiO$_2$/Si used as growth substrate. As shown in Figure 1a, the hBN film shows uniform nanoscale flatness over an area of 10x10 µm$^2$. The morphology is qualitatively comparable to that of the bare silica substrate (Figure 1b). Indeed, we retrieve an average root mean square (RMS) roughness of 450 pm for the $SiO_2$/Si substrate used as target for the growth, and of 935 pm for the hBN (see Figure S1). In inset of Figure 1a-b we report a representative AFM line profiles for commercial $SiO_2$/Si and IBAD-hBN. This result indicates that the IBAD growth process maintains the surface morphology in a range potentially suitable for high-quality graphene-based devices, even for thick hBN films. A very low roughness is in fact instrumental for a material to be used as graphene substrate, since local strain variations are regarded as a major source of carrier scattering in graphene[41].

In Figure 1c we shown the angle dependent study of the x-ray absorption near edge spectroscopy (XANES) for our hBN: around 192 eV we have the energy transition from *1s* to *π\** for boron [32]; the peak intensity in this transition follow a cosine square dependence with the incident angle[42] suggesting that the hBN crystals have a preferential orientation parallel to the substrate plane.

In Figure S1c we report a representative Raman spectrum of the synthesized hBN film. We observe two main Raman modes: the one at ~ 1370 $cm^{-1}$ is attributed the characteristic $E_{2g}$ vibrational peak of hBN, while the Si third order transverse optical (3$^{rd}$ TTO) peak [43] is located at ~ 1450 $cm^{-1}$. The full-width-at-half-maximum (FWHM) of the $E_{2g}$(hBN) peak, an indication of the material crystallinity, is 37 $cm^{-1}$, higher than that measured for single crystal exfoliated hBN (~8 $cm^{-1}$) [44], but comparable to that reported for CVD-grown hBN[45].

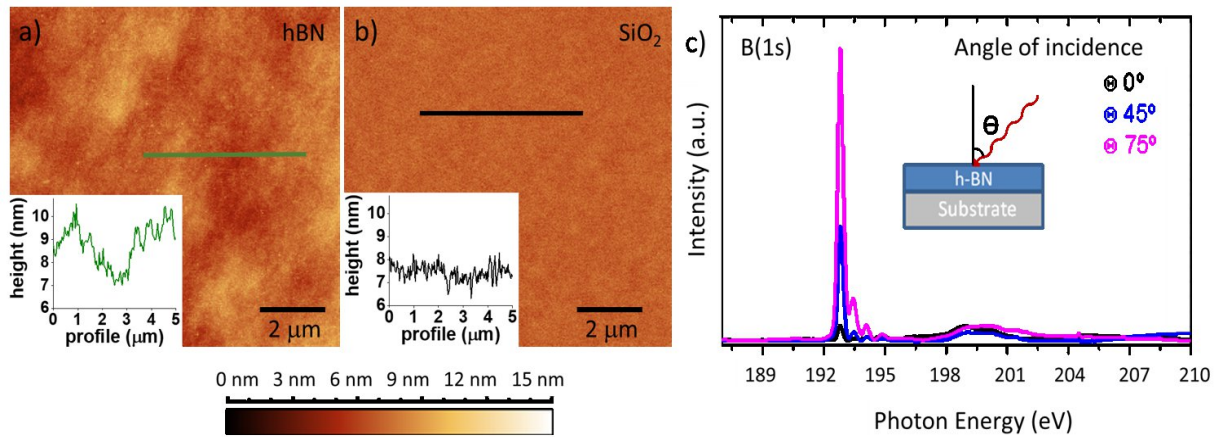

*Figure 1 AFM micrograph of (a) hBN, and (b) $SiO_2$/Si over an area of 10x10 µm$^2$. The color map range for both images is 0-15 nm. In inset: representative AFM line profiles of $SiO_2$/Si (black) and hBN (green). c) B(1s) angular XANES from hBN to determine the orientation of basal planes.*

Figure 2a reports representative Raman spectra for graphene single-crystals transferred on $SiO_2$ (black) and on hBN (green). The characteristic graphene 2D and G Raman peaks are observed around ~2675 cm$^{-1}$ and 1582 cm$^{-1}$, respectively, while the D-peak (~1350 cm$^{-1}$) is absent, indicating that defects are negligible[46]. The 2D-peak can be fitted with a single Lorentzian, as expected for monolayer graphene [47], with comparable FWHM values averaging at 25 cm$^{-1}$ and 23 cm$^{-1}$ on SiO$_2$ and hBN, respectively, suggesting a low amount of strain fluctuations (Figure 2b). The FWHM of the G-peak is found to average at ~12 cm$^{-1}$ and 10.5 cm$^{-1}$ for SiO$_2$ and hBN, respectively, as shown in Figure 2c. The A(2D)/A(G) values, reported in Figure 2f, suggest a carrier concentration within the intrinsic limit for graphene on hBN and close to 100 meV for graphene on SiO$_2$[48]. Also, the increased I(2D)/I(G) value for graphene on hBN (Figure 2e) indicates a reduction of the doping level. In Figure 2d we report the correlation plot between the 2D and G peak position[49][50]. Although the data collected for graphene on SiO$_2$/Si present a narrower dispersion than those on hBN, compatible with the higher roughness of the hBN measured from the AFM, both are indicative of slight compressive strain.

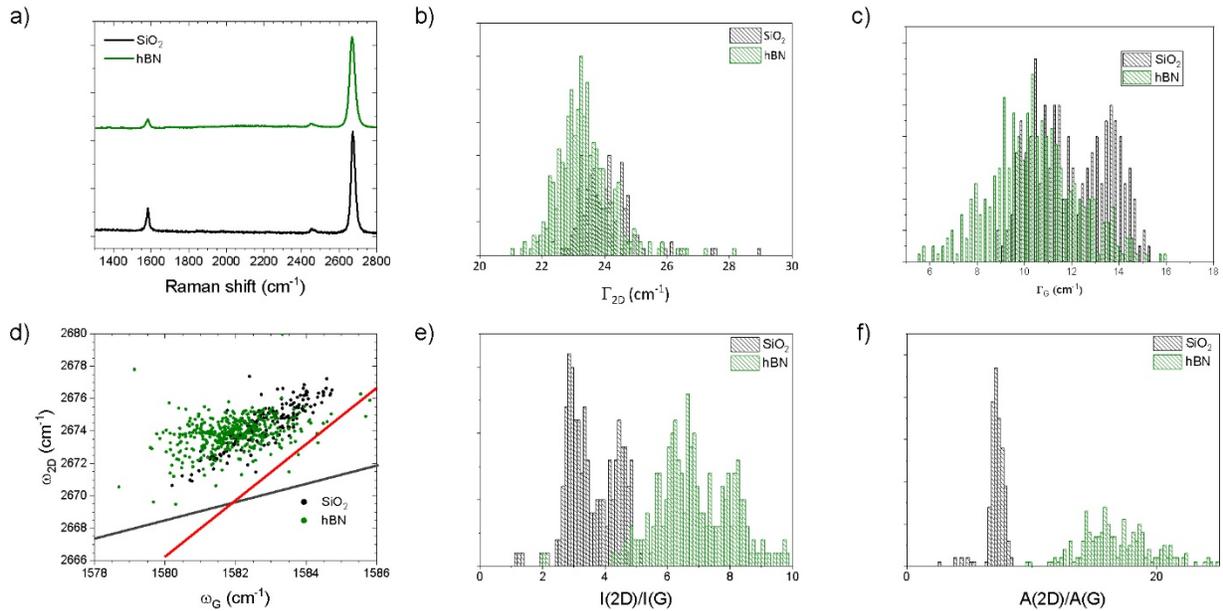

Figure 2 a) Representative Raman spectra of graphene transferred on SiO$_2$ (black) and IBAD-hBN (green). b) Distribution of the 2D-peak FWHM on the two substrates. c) Distribution of the G-peak FWHM.. d) Correlation plot of 2D-peak position as a function of the G-peak position. In d), we show as reference the dependence on strain for un-doped graphene (red, according to the Grüneisen parameter[51]), as well as the dependence on doping for the un-strained case (gray). e) Histogram of the distribution of the 2D/G peak intensity ratio and f) distribution of the A(2G)/A(G).

Figure 3a shows a sketch of the typical graphene field effect transistors (g-FETs) fabricated to investigate the transport properties of graphene when transferred on top of the IBAD-hBN. Figure 3b shows a representative transfer curve for a graphene/hBN device: employing the constant-mobility model ($R = \frac{L/W}{en\mu}$) for this device we obtained mobility of $\mu_e$ = 9500 cm$^2$/Vs and $\mu_h$ = 10400 cm$^2$/Vs for electron and holes, respectively. Those values represent an increase of ~30% compared to the same graphene crystals on SiO$_2$[52][40]. In Figure 3c we show the carrier concentration *n* as function of the back-gate voltage measured from the Hall effect: the obtained values are in line, within a 5% of error, with the expected value $n = \frac{(V_{Gate}-V_{Dirac})\varepsilon\varepsilon_0}{et}$ for a 275 nm thick SiO$_2$ plus 10 nm thick hBN, used in the constant-mobility model. The increase in the carrier mobility correlates with a reduction of the residual carrier density close to charge neutrality, as reported in Figure 3e: n$^*$ were retrieved to be 2x10$^{11}$ cm$^{-2}$ and 3.8x10$^{11}$ cm$^{-2}$ for graphene on hBN and SiO$_2$, respectively. Although the morphology of the IBAD-hBN is slightly rougher than SiO$_2$, these results indicate that a higher-quality potential landscape for the graphene carriers is provided by the nanocrystalline substrate. Furthermore, the Dirac point for this device is retrieved at 5 V, which corresponds to a charge density of 3.8 10$^{11}$ cm$^{-2}$, in line with the Raman estimation. Overall, both Raman spectroscopy[41] and electrical measurements indicate a reduced doping and residual carrier density for graphene on hBN (see Figure 2b, Figure 2e and 3e), which explain the improved transport properties measured in the graphene/hBN heterostack.

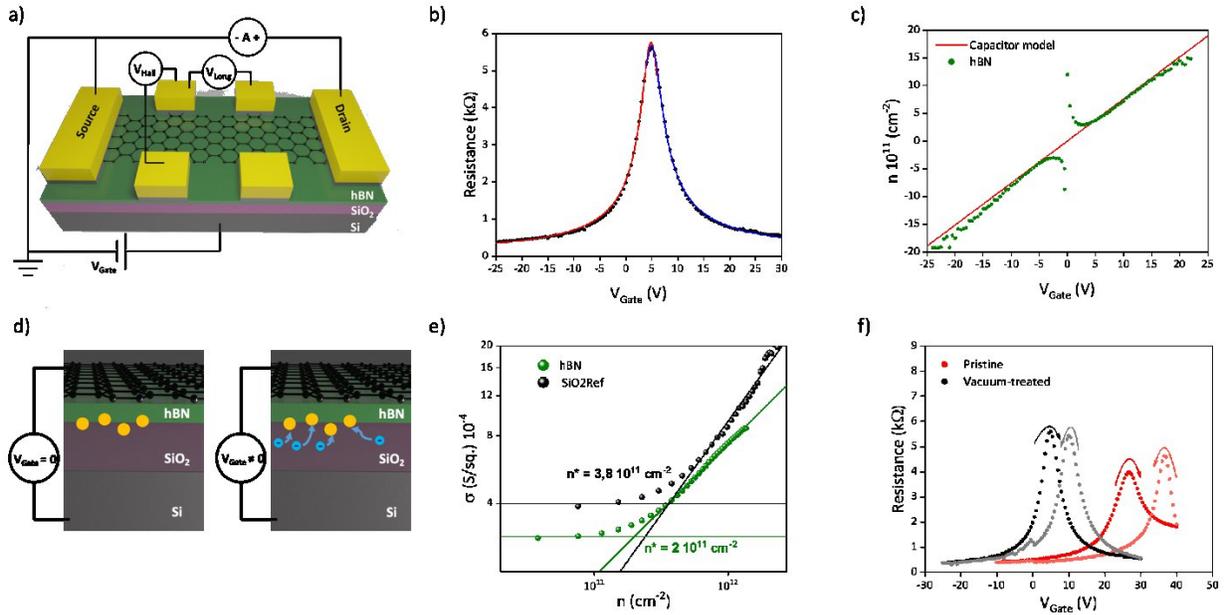

*Figure 3 a) Schematic representation of a g-FET device and measurement scheme. b) Representative transfer curve of a graphene/hBN device: from constant-mobility model we obtained mobility of 9500 and 10400 $cm^2/Vs$ for electrons and holes, respectively. c) carrier concentration on graphene as function of the backgate voltage: the values obtained from a direct Hall measure (green dots) are in good agreement with those expected for a parallel-plate capacitor with dielectric of 285 nm thickness and 3.9 dielectric constant (red continuous line). d) Schematic representation of the charge trapping in the hBN/$SiO_2$ interface, affecting the actual field effect on graphene. e) Double log plot of conductivity as a function of carrier concentration. The intersection of the minimum conductivity (horizontal lines) and a linear fit to ln(σ) vs ln(n) determine the value of the residual carrier density $n^*$, for graphene on SiO2 (black) and hBN (green). f) Transfer curve of a g-FET immediately after fabrication (red) and after 4 months in vacuum (black): the initial hysteresis largely reduced, and the carrier mobility is kept 30% higher than what was tested on $SiO_2$/ Si substrate.*

It should be mentioned that the device above was measured after keeping the structure in static vacuum (~10 mbar) for prolonged time (>4 months) after fabrication. When measured immediately after fabrication, the devices presented a pronounced hysteresis of the transfer, as shown in red curve in Figure 3f, while after storage in vacuum the hysteresis was strongly reduced, as shown by the black curve in Figure 3f. Concurrently, we also observed a shift in the Dirac point to lower doping values (i.e., $V_{Gate}$<10 V). Also, no significant variation in the carrier mobility was observed after vacuum storing, as reported in Figure S 4. All the electrical measurements reported have been performed in ambient (not vacuum) conditions, and it should be mentioned that there was no reappearance of hysteretic behavior in the sample after vacuum storage. The gate hysteresis can be attributed to charge traps that partially screen the back-gate potential. These traps can be present either at the SiO$_2$/hBN or at the hBN/graphene interfaces. We report that these traps are characterized by slow charging and discharging times, as we observe different transfer curve behaviors for different gate sweeping rates (Figure S3). We were able to induce reduction of the hysteresis also by annealing the sample at 130 °C overnight in high vacuum (10$^{-9}$ mbar). However, annealing at higher temperature for shorter times affected the transport properties of the device with a consequent mobility reduction of ~50% (see Figure S5). Moreover, the carrier concentration in the graphene shows a linear dependence on the gate voltage (see Figure 3c), with no sign of direct charge transfer.

To further assess the location of the charge traps we performed transport measurements in a top-gated exfoliated hBN/graphene/IBAD-hBN/SiO$_2$ heterostructure. When using the top-gate (that is, applying the gate potential through the exfoliated hBN flake), we did not observe significant hysteresis, nor gate sweeping speed dependence. The electrical behaviour of the fully-encapsulated device was found to be qualitatively compatible with that measured for similar devices fabricated on SiO$_2$/Si substrates (see Figures S7b and S7d): the fully-encapsulated graphene presented a significantly higher carrier mobility of ~ 15000 cm$^2$/Vs (at 5x10$^{11}$ cm$^{-2}$) and lower residual carrier density of n$^*$ = 8.5x10$^{10}$ cm$^{-2}$, as the top hBN protects from environmental contaminations. Instead, measuring the same device in back-gated configuration led to the observation of a large hysteresis (see Figure S7c) comparable to that reported by the red curves in Figure 3f. This confirms that the charge traps are present at the remote hBN/SiO$_2$ interface, likely forming during the IBAD growth process. In Figure 3d we show a representative schematic of the expected effect that charge impurities present at the interface hBN/SiO$_2$ may have on the transfer curve: when a gate potential is applied these impurities can act as traps for the electrons, leading to a non-linear change of the electric field on the graphene with the applied gate potential. Vacuum storage appears effective in removing such traps, hence ultimately eliminating hysteretic effects.

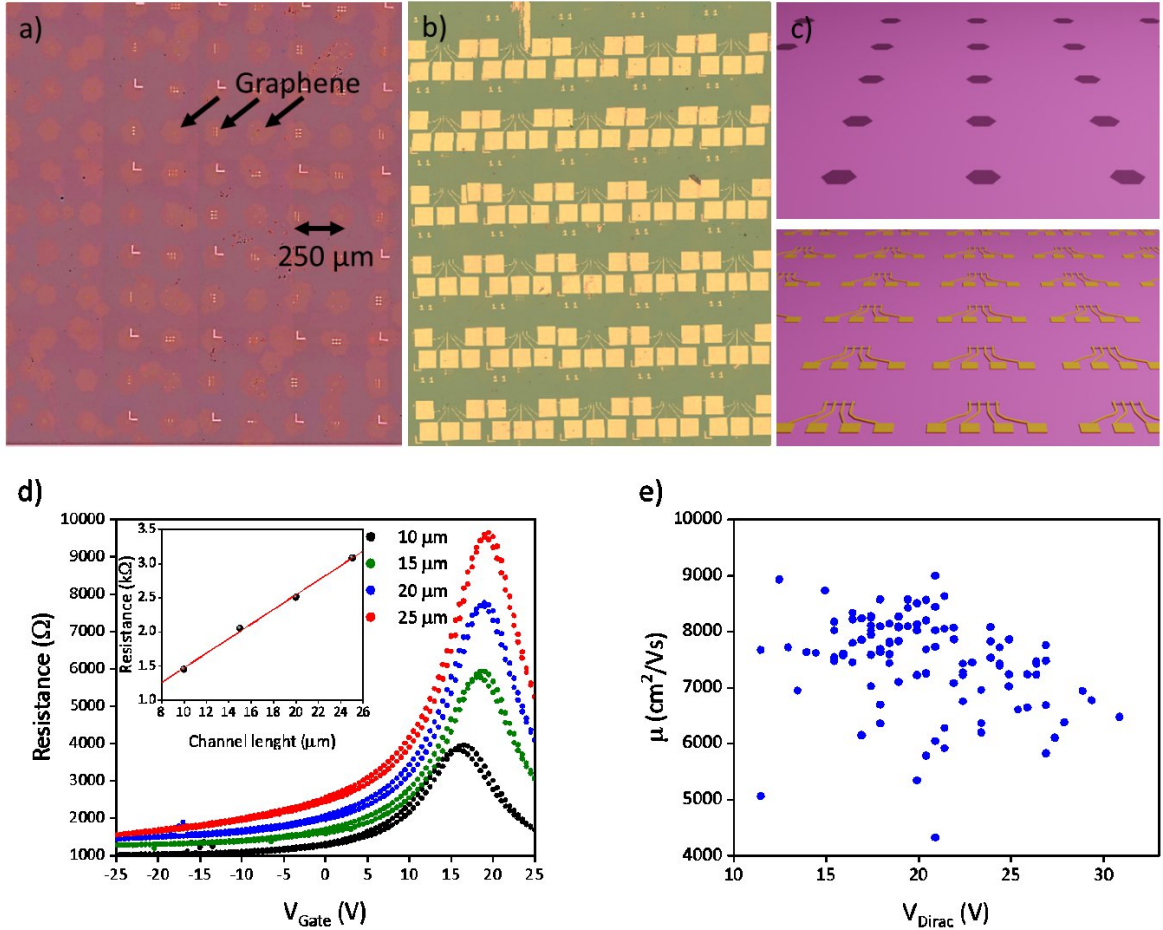

*Figure 4 a) Optical image of a seeded single-crystal graphene array deterministically transferred on hBN substrate. b) Optical image of an array of TLM devices, the total amount of devices tested is 109. c) Schematic representation of the device fabricated on the graphene array deterministically transferred on a IBAD-hBN substrate, as described in methods; the regular graphene pattern allows to realize a matrix of identical devices with constant spacing. d) Representative transfer curve of the four channels of a TLM device. In the inset, we report the resistance measured at fixed carrier concentration ($10^{12}$ cm$^{-2}$) as a function of the channel length for a representative device. By the linear fit as a function of the length, it is possible to isolate the contribution of the contact and channel resistance, and then obtain the graphene mobility* [53]. *e) Distribution of the measured mobility as a function of the position of the Dirac point.*

Finally, to prove the scalability of our approach we realized and electrically characterized more than 100 graphene devices, transferred on IBAD-hBN. As previously reported[33], we have developed approaches to grow single-crystal graphene in a deterministic pattern and to precisely transfer such matrixes on a desired substrate; thus, we can scale-up the statistics on the device number by realizing a matrix of TLM devices. Each TLM device was realized on a different graphene single-crystal of the array, as shown in Figure 4a-c. Fabrication of each TLM channel

(width 10 μm; length varying from 10 to 25 μm, in 5 μm steps) within the same graphene crystal allowed us to isolate the contact and channel contribution to the resistance, and thus estimate the graphene mobility[53]. The devices are realized with the same orientation, which means appreciatively the same crystallographic orientation, as previously reported[54]; however, the electrical behavior of the devices should not be influenced by the crystallographic orientation in our regime of measure. In Figure 4d are reported the transfer curves for a representative TLM device. In Figure 4e we report a plot of the carrier mobility as a function of the gate voltage position of the Dirac point. The mobility obtained from the constant mobility model in the array, was found to be as high as 9000 $cm^2$/Vs, with average μ at ~ 7500 ± 850 $cm^2$/Vs. The TLM sample was subjected to a shorter (few weeks) vacuum treatment with respect to the gFET and for this reason the average Dirac point is found to be significantly higher than expected, i.e., 20.5 ± 4 V, while the hysteresis is still reduced. The mobility values reported indicate a substantial improvement with respect to large-scale characterization of the same CVD graphene crystals on $SiO_2$ (average μ ~ 5000 $cm^2$/Vs)[55], in line with the 30% improvement reported for the gFET devices. The results were obtained on a 1x1 $cm^2$ chip, but could be straightforwardly extended to wafer-scale via multiple tile transfer of the graphene matrixes[55].

## 3. Conclusions

In summary, we have realized and characterized scalable vertical hBN/graphene heterostructures which allow for the realization of devices with promising electronic performances. Nanocrystalline hBN was grown via IBAD on $SiO_2$/Si, and presents a thickness of 10 nanometers, which makes it suitable to be used both as an encapsulant and as a gate dielectric. Microscopic characterization was performed to investigate the surface morphology of the scalable hBN, which was found to be comparable in terms of flatness to those of the $SiO_2$ growth substrate. Spectroscopic and transport measurements were carried out to compare the properties of graphene when transferred on commercial $SiO_2$/Si and hBN. We observe a relevant improvement in terms of residual carrier density and carrier mobility, that indicates how the adopted hBN provides a high-quality landscape for the graphene carriers. Also, we demonstrate that the hysteretic behaviour observed in heterostructure realized with as-received hBN can be significantly reduced by vacuum treatment of the material. To prove the scalability of our approach we tested an array of graphene crystals, transferred on IBAD-hBN over centimeter scale, and obtained reproducible mobility values exceeding 7500 $cm^2$/Vs. Future developments might concern a deeper investigation of the charge trapping mechanism at the hBN/SiO$_2$ interface, as well as the optimization of the IBAD-hBN transfer to realize fully-encapsulated and scalable hBN/graphene/hBN heterostructures.

## Acknowledgments

The research leading to these results has received funding from the European Union's Horizon 2020 research and innovation program under grant agreement no. 881603-Graphene Core3.

# Supporting information
# Scalable high-mobility graphene/hBN heterostructures

*Leonardo Martini[1,*], Vaidotas Miseikis[1,2], David Esteban[3], Jon Azpeitia[3], Sergio Pezzini[4], Paolo Paletti[1,2], Michal W. Ochapski[1,2], Domenica Convertino[1], Mar Garcia Hernandez[3], Ignacio Jimenez[3], Camilla Coletti[1,2,*]*

1. Center for Nanotechnology Innovation @NEST, Istituto Italiano di Tecnologia, Piazza San Silvestro 12, 56127, Pisa, Italy
2. Graphene Labs, Istituto Italiano di Tecnologia, Via Morego 30, I-16163 Genova, Italy
3. Instituto de Ciencia de Materiales de Madrid, Consejo Superior de Investigaciones Científicas, E-28049 Madrid, Spain
4. NEST, Istituto Nanoscienze-CNR and Scuola Normale Superiore, Piazza San Silvestro 12, 56127, Pisa, Italy

*leonardo.martini@iit.it, camilla.coletti@iit.it*

a) BN and BNC films

BNC films with a thickness of 10 nm and with crystalline orientation parallel to that of the target substrate were synthesized using $B_4C$ as precursor and their morphology and spectroscopic features were compared to those of the BN films. AFM analysis, reported in Figure S1, reveals a rougher morphology for the BNC films (root mean square (RMS) roughness of 950 pm) with respect to BN films of the same thickness (RMS roughness 935 nm). Raman spectroscopy indicates that both BNC films present a much lower $E_{2g}$ Raman peak, compared to the BN film, suggesting a lower crystalline quality.

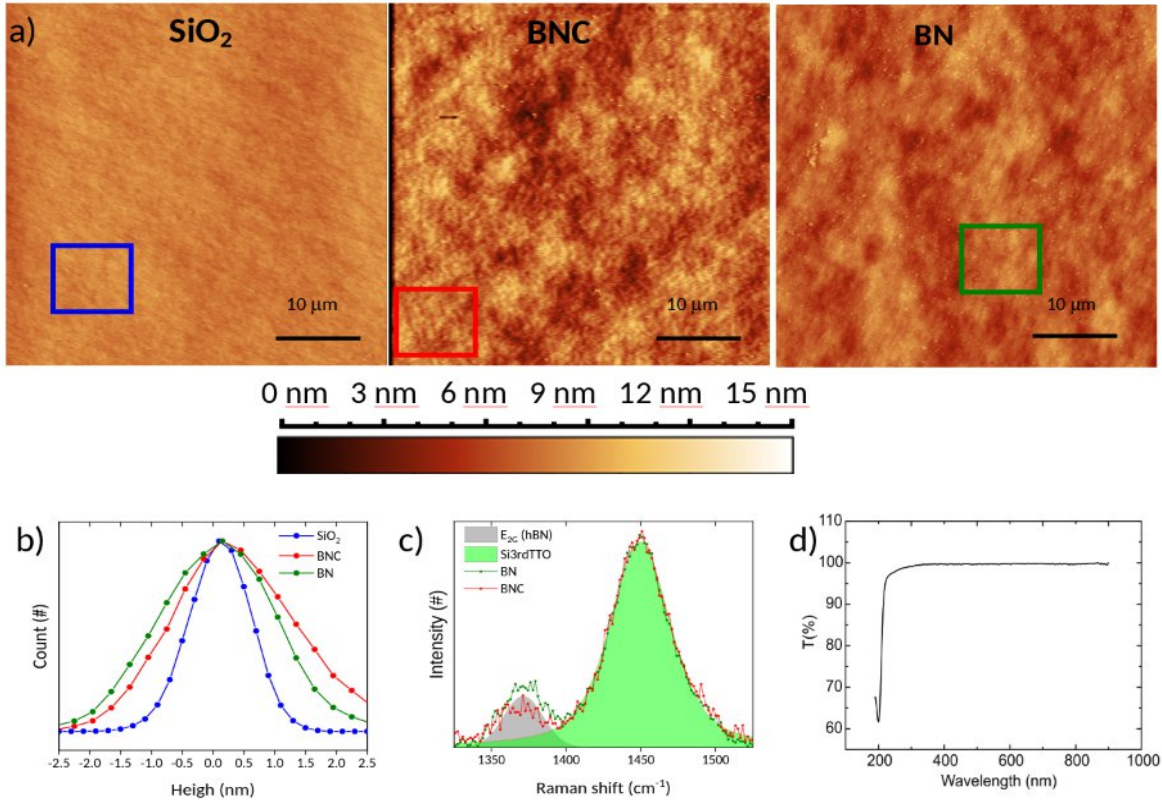

*Figure S 1 a) AFM micrographs of representative BN and BCN samples, compared with the bare SiO$_2$/Si growth substrate. b) Height distribution of the different substrates, over a 10x10 µm$^2$ area. c) Representative Raman spectra for the BN and BNC samples. Both samples show a E$_{2g}$(hBN) peak at ~1370 cm$^{-1}$, with similar FWHM ~ 37 cm$^{-1}$. The E$_{2g}$(hBN) peak shows lower intensity for the BNC sample. d) UV-Vis spectrum of hBN grown on sapphire. The hBN thickness is determined applying the Lambert Beer relation ($I = I_0 e^{-\alpha d}$, where $I_0$ is the initial intensity, $I$ is the intensity at 200 nm, α is the absorption coefficient at 200 nm and d is the thickness of the film)*

The FWHM of the 2D Raman peak of graphene single-crystals transferred on SiO$_2$ and BN are comparable (i.e., 25 and 23 cm$^{-1}$, respectively), while larger values are retrieved for BNC substrates (i.e., 30 cm$^{-1}$). Raman correlation plots suggest that graphene doping and strain are comparable on BN and SiO$_2$, while BNC performs worse as a substrate.

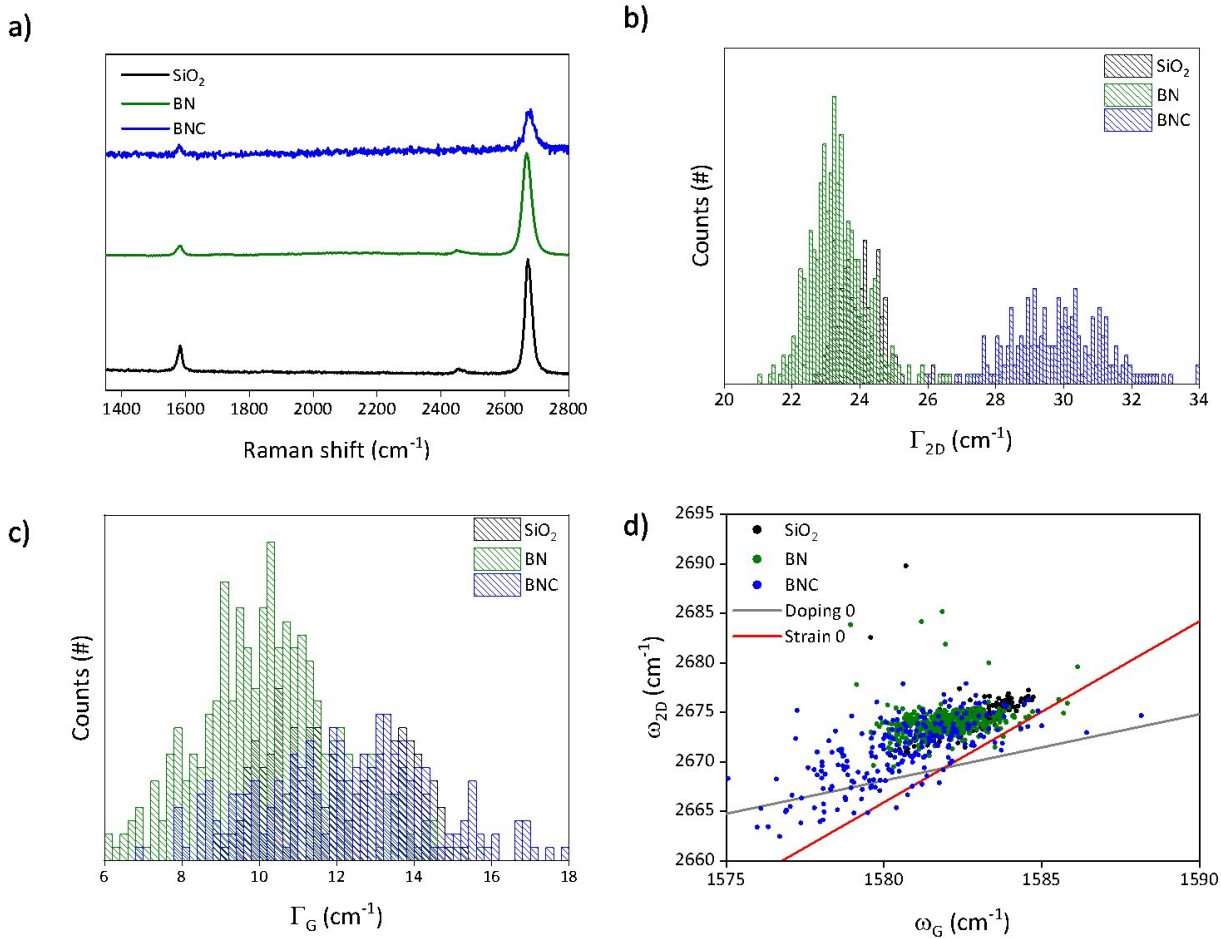

*Figure S 2 a) Comparison of representative Raman spectra of graphene transferred on $SiO_2$ (black), 10nm-thick BN in green, and 10nm-thick BNC in blue. b) Distribution of the 2D-peak FWHM. The average value for the graphene transferred on $SiO_2$ and BN is of 25 and 23 $cm^{-1}$ respectively, while on BNC is of 30 $cm^{-1}$. c) The distribution of the G-peak FWHM show no relevant change between the different substrates, averaging around 13 $cm^{-1}$ for both BNC and $SiO_2$. d) Correlation plot between the position of the G-peak and 2D-peak; as reference, we show the evolution with strain for un-doped graphene (black), and with doping for the un-straied case (red).*

### b) Gate hysteresis and effect of annealing

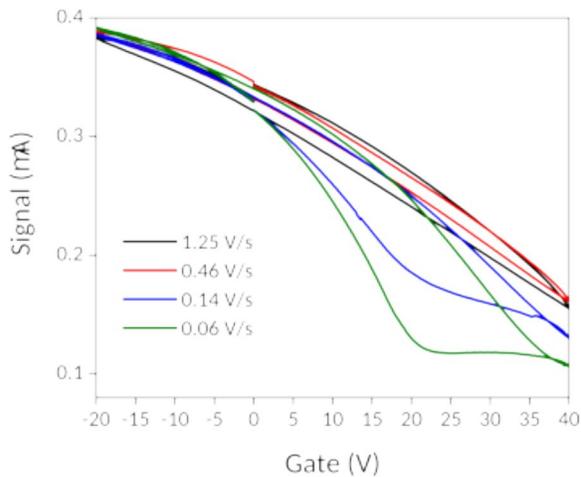

*Figure S 3 Hysteresis dependence on the sweep rate: transfer curves on the same devices, performed at different gate sweeping speeds. The hysteresis increases at slow rate, indicatihng a relatively slow charge trapping dynamic.*

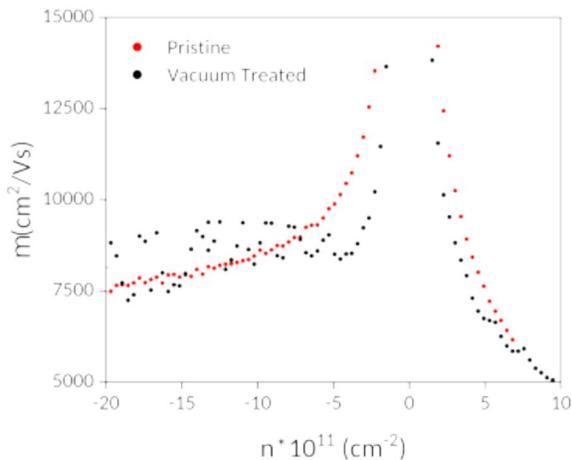

*Figure S 4 S In red we report the mobility as function of the carrier concentration, obtained from a Hall measurement: carrier mobility at technological-relevant concentration of $10^{12}$ cm$^{-2}$ are $\mu_e$ = 5000 cm$^2$/Vs and $\mu_h$ = 7500 cm$^2$/Vs for holes and electrons, respectively. The difference in the mobility with respect to the values reported in the main text is expected from the different assumption in the two methods [53]. In black we report the Hall mobility as function of the carrier concentration for the same device after the vacuum treatment: we do not observe significant changes in the mobility.*

Hysteresis in the transfer curves is generally an unwanted behavior in electronic devices, which has to be solved in order to make devices suitable for commercial applications. Keeping the prepared sample in a commercial desiccator has proven to work well, even if the time needed to reduce the hysteresis to a level comparable with graphene on SiO$_2$/Si is of the order of several weeks. To accelerate this process, we tested high-vacuum thermal annealing on the graphene/hBN devices. We annealed several samples at different temperatures and for different times: low-temperature annealing shows some promising results, with a relevant reduction in the hysteresis as shown in Figure S5a for a 130 °C process. However, this kind of annealing tends to be slow. On the opposite, annealing at higher temperature tends to degrade graphene quality, as shown by the pinning in the electron branch of the transfer curve of the sample annealed at 300 °C, as shown in Figure S5b. The degradation of graphene is also proven by Raman, as the 2D-FWHM increases from 23 cm$^{-1}$ to more than 35 cm$^{-1}$, as shown in Figure S5d. Also the 2D/G peak intensity ratio is reduced, as shown in Figure S5c.

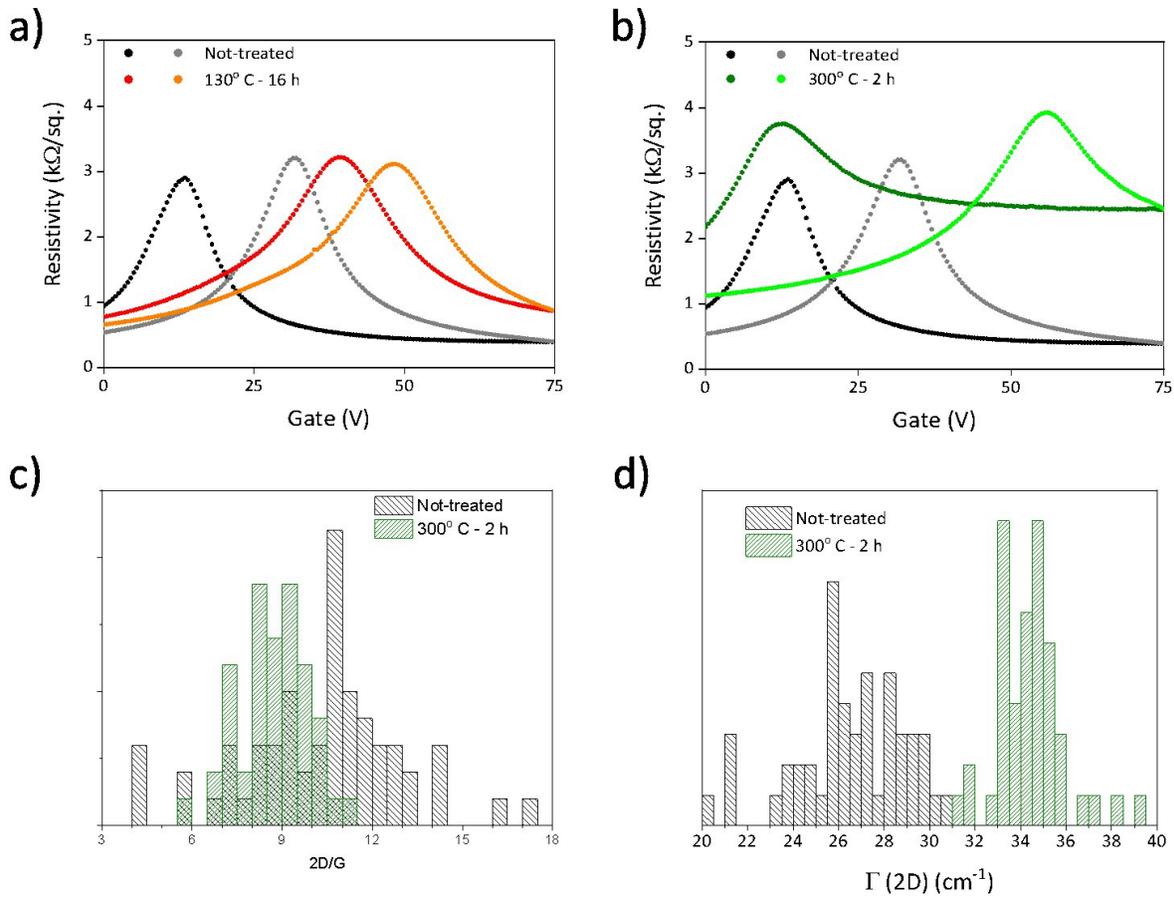

*Figure S 5 A reduction of the hysteresis can be induced via ultra-high vacuum annealing at relatively low temperature (130 °C) for 16 hours. Annealing for shorter time didn't induce any reduction of the hysteresis, while increasing further the temperature led to a degradation of graphene. a-b) Transfer curve for comparable devices after fabrication*

*and after 2 different annealing procedures: 130 °C – 16 hours in a) and 300 °C – 2 hours in b). c) 2D/G intensity ratio shows a reduction after the 300 °C annealing. d) The increase of the average 2D-FWHM after the 300 °C annealing also suggests a degradation of graphene.*

### c) Fully encapsulated graphene

To overcome the gating limitation at the untreated interface $SiO_2$/hBN in the system and prove the final achievable performance of this scalable graphene/hBN heterostructure, we implemented a top-gated device, using an exfoliated hBN flake as top-gate dielectric.

Through dry pick-up technique with exfoliated hBN, we transfer a portion of a graphene crystal[56], totally comparable with the one used in previous characterization, on IBAD-hBN. We use AFM and Raman (Figure S6) characterization to select a flat area for device fabrication. In Figure S7a we show a false-colour SEM image of the final device, we coloured in red the top-gate contact, in yellow the side contacts and in green the hBN/graphene heterostructure. The transfer curve obtained using the top-gate is reported in Figure S7b: the hysteresis is strongly reduced and the overall behaviour is compatible with similar devices made on commercially available $SiO_2$/Si substrates. The transfer curves are also more stable upon changing the gate sweeping speed.

Conversely, performing transfer curves on the same device using the back-gate (Figure S7c), we obtain the same hysteresis and overall behaviour of the fresh devices with exposed graphene. This result further indicates that the charge trapping responsible for the gate hysteresis takes place at the hBN/$SiO_2$ interface, unaffected by the dry transfer of the graphene with exfoliated hBN.

Finally, we test the overall performance of the fully encapsulated material, through Hall measurement. As shown in Figure S7d the carrier mobility reaches values >20 000 $cm^2$/Vs, remaining above 5 000 $cm^2$/Vs even at carrier concentration >$10^{12}$ $cm^{-2}$. The further improvement over the exposed graphene devices can be explained by the absence of environmental contamination on the surface of the graphene. This is also supported a reduction of the residual carrier density, reported in Figure S7d. The value of $n^*$ = $8.5\times10^{10}$ $cm^{-2}$ is half the one of the exposed sample shown Figure 3e, in agreement with the increased values of carrier mobility.

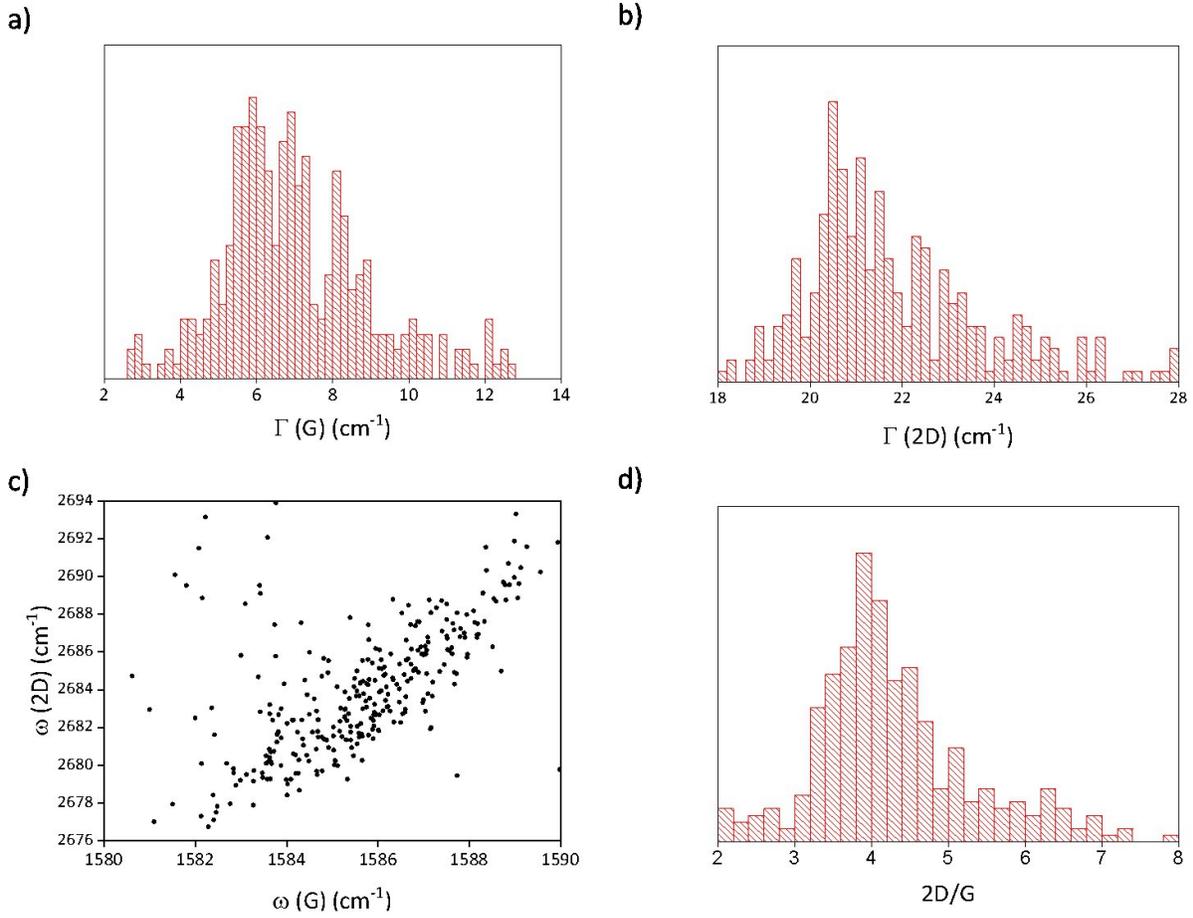

*Figure S 6 Raman characteristics of the fully-encapsulate graphene sample. The G-peak FWHM in a), the 2D/G position correlation plot in c) and 2D/G intensity ration in d), suggest low doping level. b) The 2D FWHM, lower than the exposed graphene transferred both on $SiO_2$ and BN substrate, suggest sreduced strain variations. No D-peak has been observed in all the spectra acquired.*

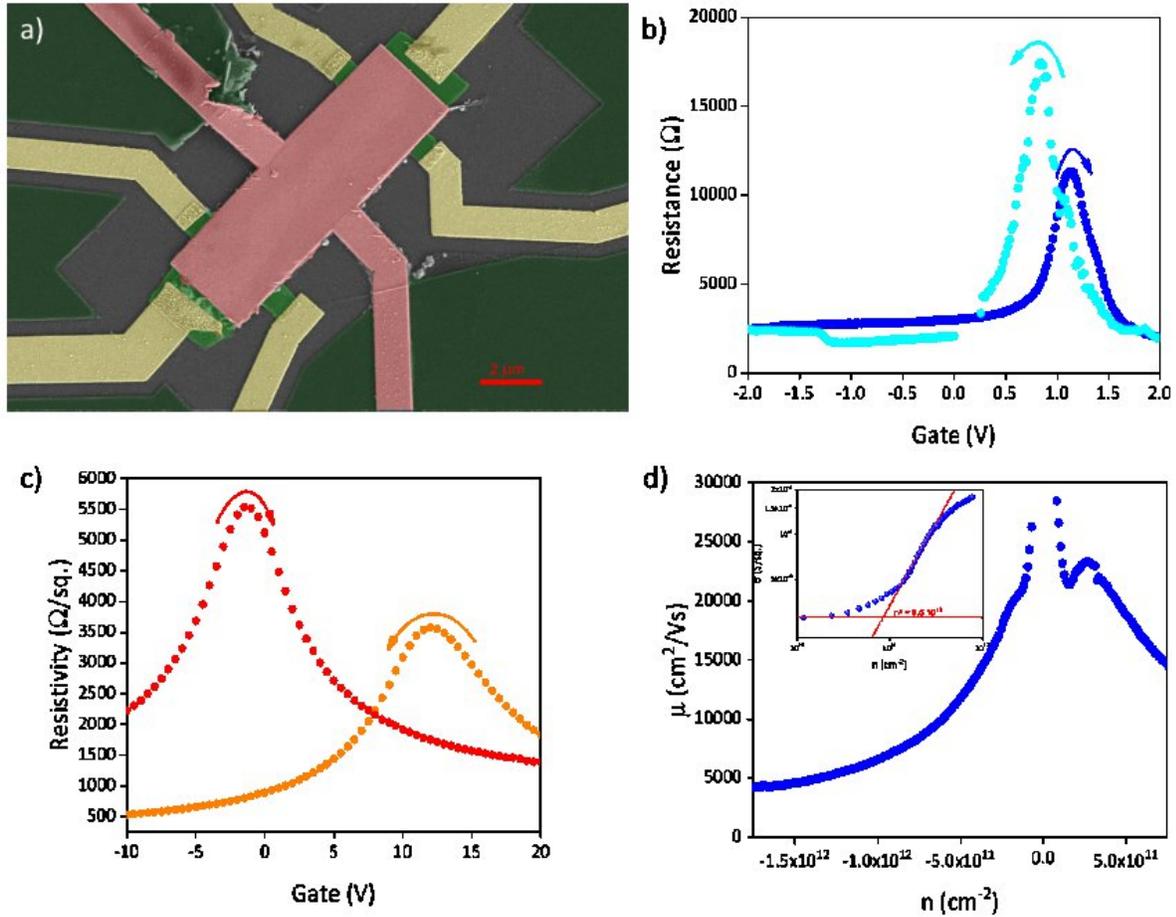

*Figure S 7 a) false-color SEM image of the fully-encapsulated device: in dark green the unetched hBN/graphene, in yellow the metal side contacts, in red the top-gate and in green the device channel. The actual channel width is 6.6 µm and total length is 33 µm; several lateral contacts are presents each couple 8.7 µm apart, conferring an aspect ratio of 1.3 to the longitudinal measurements b) Transfer curve of the graphene on IBAD-hBN, using exfoliated hBN as the dielectric in a top-gated configuration: sweep from negative to positive (light blue) and from positive to negative (blue) gate values show reduced hysteresis. c) Transfer curve of the fully-encapsulated graphene, performed using the $SiO_2$/hBN back-gate. The hysteresis and difference between forward and backward sweep are present and comparable to the exposed-graphene devices. d) Moblility as function of carrier concentration for graphene on PVD-hBN top-encapsulated with exfoliated hBN. Inset: the double log plot of σ versus n indicates a residual carrier density of $n^* = 8.5 \times 10^{10}$ $cm^{-2}$.*